\documentclass{jpsj3}

\newcommand{\argmin}{\mathop{\rm arg~min}\limits}

\usepackage{txfonts}
\usepackage{here}

\title{
       Virtual Screening of Chemical Space based on Quantum Annealing
       }

\author{
Takuro Tanaka$^1$\thanks{takuro.tanaka@lgjlab.com}, Masami Sako$^1$, Mahito Chiba$^1$, 
Chul Lee$^2$, Hyukgeun Cha$^2$, and Masayuki Ohzeki$^{3,4,5}$
}
\inst{
$^1$ LG Japan Lab Ltd., Yokohama, Kanagawawa, Japan \\
$^2$ LG Electronics Inc.,19, Yangjae-daero 11gil, Seocho-gu, Seoul 137-130, Korea \\
$^3$ Graduate School of Information Sciences, Tohoku University, Sendai, Japan \\
$^4$ Department of Physics, Tokyo Institute of Technology, Tokyo, Japan \\ 
$^5$ Sigma-i Co. Ltd., Tokyo, Japan \\
}

\abst{ 
For searching a new chemical material which satisfies the target characteristic value, for example emission wavelength,  
many cut and trial of experiments/calculations are required since the chemical space is astronomically  large (organic molecules  generates  $>10^{60}$  candidates).
Extracting feature importance is a method to reduce the chemical space, and limiting the search space to those features leads to shorter development time.
Quantum computer can generate sampling data faster than classical computers, and this property is utilized to extract feature importance.
In this paper, quantum annealer was used as a sampler to make data for extracting feature importance of material properties.
By screening the chemical space with feature importance,
it was found that the chemical space can be reduced to less than 1 percent.
This result suggests that the acceleration of material research can be achievable.
}

\begin{document}
\maketitle


 Quantum computer is expected as a key technology to change from conventional computer.
Based on quantum annealing, a combinatorial optimization from wide search range can be obtained.  
The quantum annealer is sometimes regarded as a simulator for the quantum many-body dynamics \cite{ref1,ref2}. 
Practical applications of quantum annealer have been presented across various fields, such as finance \cite{ref3,ref4,ref5}, 
traffic \cite{ref6,ref7}, routing optimization\cite{Volkswagon, Ohzeki, Haba},
logistics \cite{ref8,ref9}, manufacturing \cite{ref10}, and marketing \cite{ref11}, 
as well as in decoding problems \cite{ref12,ref13}. 

Its potential for solving the optimization problem with inequality constraints has been enhanced \cite{ref14}, 
especially in the case that is hard to formulate directly \cite{ref15}. The comparative study of quantum annealer has 
also been performed for benchmark tests to solve the optimization problems \cite{ref16}. 
The quantum effect on the case with multiple optimal solutions has also been discussed \cite{ref17,ref18}. 
Further, applications of quantum annealing for machine learning for solving optimization problems have been reported \cite{ref19,ref20,ref21,ref22,ref23,ref24}.
As in the case for material informatics (MI) in the classical computer,
optimization scheme based on quantum annealing can be very useful approach.

Recently, with the need for higher performance and diversification of materials, it is indispensable to accelerate research and develop new materials with unprecedented properties and functions.
However, since the properties of materials depend on the many microscopic elements, one needs examination of huge combinations in chemical space.
To shorten development time, therefore, one needs algorithm to search materials that satisfy the desired properties with few trail.

In previous study, combinatorial optimization problem is applied to material research \cite{Oyaizu,Tamura}.
Hatakeyama et al. \cite{Oyaizu} extracted high-melting-temperature molecules using an open experimental database (DB) of organic molecules \cite{Oyaizu-ref}. 
Kitai et al. \cite{Tamura} studied on designing complex thermofunctional metamaterials consisting of SiO2, SiC, and Poly(methyl methacrylate). 
In their study, the DB was constructed using atomistic simulation to calculate the performance for thermal radiator.

 Experiments on D-Wave quantum annealer hardware platforms have demonstrated that the quantum annealing hardware behaves like a Gibbs-Boltzmann sampler at a hardware-specific effective temperature \cite{Sampling_LosAlamos}.
Using quantum annealer, it is possible to perform sampling according to the Gibbs-Boltzmann distribution.
When this characteristics is used effectively, it may help as means of solving Boltzmann machine and probability distribution-based optimized problems. 
However, to the best of our knowledge, there are no examples of MI problems that have been used quantum annealer as a sampling machine.

In order to search luminescence material based on quantum chemistry, the energy value of HOMO-LUMO gap is  indispensable in the DB.
In this paper,  we use QM9 as a database and extract feature importance of emission wavelength by making a sampling data based on quantum annealer.
Here we note that emission wavelength can be exchanged to HOMO-LUMO Gap in microscopic feature.
QM9 dataset was published in 2014 by Ramakrishnan et. al. \cite{QM9_1,QM9_2} consists of more than 133k organic molecules with up to 9 heavy atoms (C, N, O and F) 
with corresponding geometries, thermodynamic and electronic properties, i.e. HOMO-LUMO gap, and simplified molecular-input line-entry system (SMILES).
Note that SMILES can be converted to fingerprints, which are a generally well-known essential cheminformatics tools for mapping chemical space.
The prediction model is constructed based on Stochastic Gradient Descent (SGD) Regressor with fingerprint as a descriptor.
The cost function $y$ is given as the distance between the HOMO-LUMO gap $\Delta_{DB}$value in DB and the target HOMO-LUMO gap value $\Delta^{\ast}$ as follows:

\begin{equation} 
\label{eq0}
\begin{split} 
y = (\Delta_{DB}- \Delta^{\ast})^2.
\end{split} 
\end{equation}
Using quantum annealer, the optimized combination of fingerprint which minimize the cost function and the optimized value is obtained.
Even if there are many discriptors, annealer searches and samples various combinations probability
to obtain the combination with lowest energy.
Analysis of feature importance can be easily performed by utilizing the sampling results.
By extracting the dominant features, as a result, dimension reduction of chemical space is possible.



Molecular fingerprints encode molecular structure in a series of binary digits (bits) 
that represent the presence or absence of particular substructures in the molecule.
Typical fingerprints for cheminformatics are 
Morgan, extended circular finger print (ECFP), molecular access system (MACCS) key, Avalon, etc.
This fingerprints for molecular structure correspond to the spin state $S_z$ up ($S_z = 1$) and down ($S_z = -1$) in Ising model.
In this paper, we used Avalon 512 bit using a free library, RDKit. \cite{rdkit}
Comparing fingerprints allow us to determine the similarity between two molecules, 
to find matches to a query substructure, etc.
To extract the feature importance of material properties using a quantum annealer,
in this paper,  we first constructed an Ising type prediction model  and confirmed that the prediction accuracy is high.
Then, we verified the accuracy of the optimal solution obtained.

To predict target HOMO-LUMO gap,
the cost function $y$ is given as the distance from the target HOMO-LUMO gap: $y= f_{\rm{pred}} $ as given in Eq. (\ref{eq0}).
Here, $\Delta^{\ast}$ is set to $0.32$eV in this paper.
We define the Ising (quadratic) model as the prediction model of the target HOMO-LUMO gap as follows:
\begin{equation} 
\label{eq1}
\begin{split} 
f_{pred} = \sum_{i \neq j} Q_{ij} x_i x_j + \sum_i h_i x_i .
\end{split} 
\end{equation}
Here, descriptor $x_i$ represents fingerprint: $x_i = 0$ or $1$.
To make regression of the prediction model for the training data, we redefine the quadratic model as a linear model by using a new expression $X_{ij} = x_i x_j$.
The quadratic model can be regarded as the linear model $f_{pred} = \sum_{ij} Q_{ij} X_{ij}$, where $Q_{ii} = h_i$.
We then perform the standard regression technique for the linear model.

In terms of prediction accuracy, a prediction model using full fingerprint is appropriate. 
However, from the viewpoint of memory, it is very difficult to perform for following two reasons:
1) Since cost function is given as a quadratic term, ${}_nC_2$ interactions ($n \geq 10^2 $) must be added which makes difficult to calculate for regression.
2) The number of data is over $10^5$ for QM9 databases. 
Therefore, compressing fingerprint is indispensable.
The threshold to compress fingerprint is set as 15000, which means that fingerprint is compressed when the number of times used as 1 is less than 15000 in train data, i.e.  fingerprint less than 14\% ( $\approx 15000/110000 $).
Then, the 512 Avalon fingerprint is compressed to 207 fingerprint. 

\begin{figure}[hbtp]
\begin{center}
\includegraphics[width=.85\linewidth]{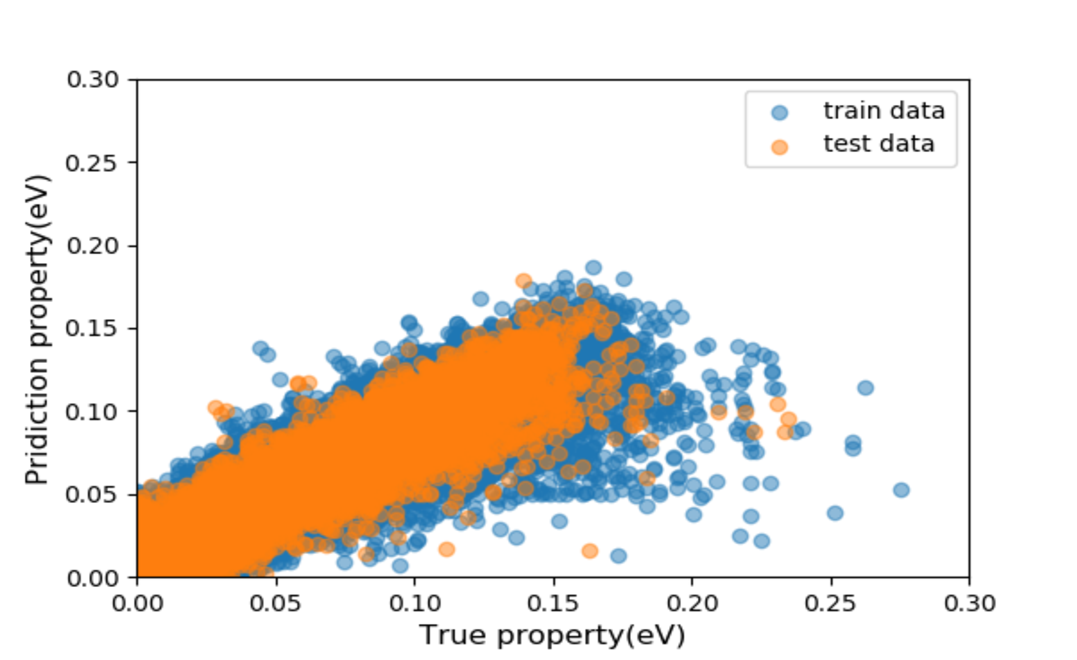}
\vspace*{-12.0cm}
\caption{
(Color online)
Regression  result using Avalon fingerprint. The  DB  was split  into training  (90\%)  and  testing (10\%)  datasets,
for  which the  $R^2$  scores  were  0.82 for train data and 0.81 for validation,  respectively.  
Here, the  dimension  of  discriptor  is set as 207.
}
\label{f1}
\end{center}
\end{figure}

\begin{figure}[H]
\begin{flushleft} %
\centering
\includegraphics[width=.85\linewidth]{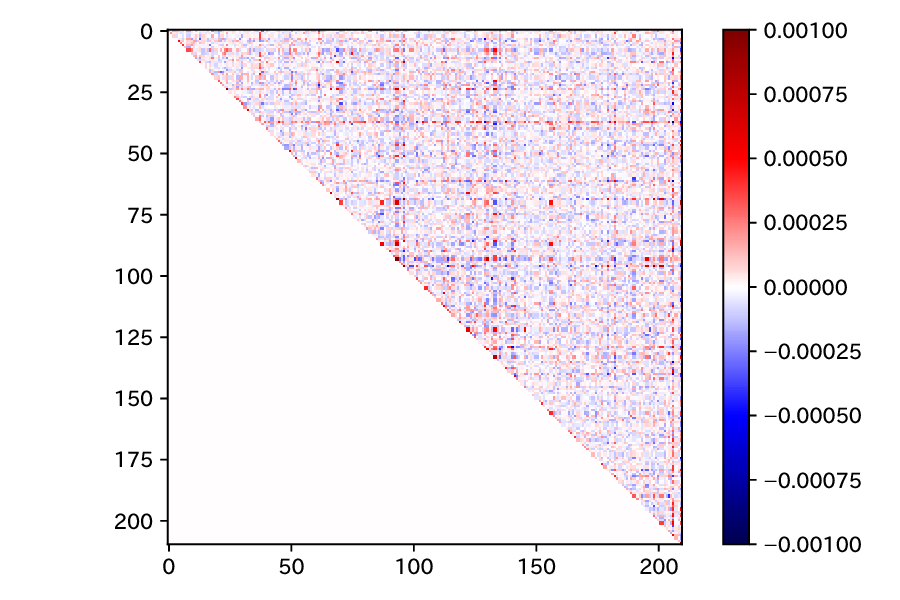}
\vspace*{-8.0cm}
\caption{
(Color online)
Coefficients in linear-quadratic form obtained by optimizing the prediction model QUBO.
}
\label{fQUBO}
\end{flushleft} %
\end{figure}

For regression, the ratio between train and test data is set to 9:1.
To make prediction model in this paper,  SGD Regressor is used which gives high accuracy model for data more than $10^5$. 
First regression with squared loss and a $L_1$ norm regularization
is calculated with the parameter of regulation strength $\alpha$ = 0.1, learning rate $\eta$ = 0.001.
In this case the loss function of SGD regression does not decrease from 0.000382 after few interation, and prediction accuracy (Coefficient of Determination $R^2$) is negative.
This results suggest that it is difficult to predict cost function $y$ from descriptor $x$ due to lack of learning.
Next, $L_2$ norm regularization with same parameter ($\alpha$=0.1, $\eta$ = 0.001) are calculated.
The loss function decreased as 0.000073 after iteration of 160 times, and  
prediction accuracy (Coefficient of Determination $R^2$) is 0.82(0.81) for train data (test data) as shown in Fig \ref{f1}. 
This result shows that with $L_2$ norm regularization, learning works satisfactorily and able to predict with a sufficient accuracy.
For the annealing calculation explained below, the coefficient of quadratic term (QUBO) with $L_2$ norm regularization is used which is shown in Fig\ref{fQUBO}.


In this study, a simulator of quantum annealing which is called openjij \cite{Jij} is used to obtain the optimized combination of fingerprint in Eq.(\ref{eq3}).
Openjij is an open-source library for heuristic optimization problems in Python 
which is based on the quantum Monte Carlo method via Suzuki-Trotter decomposition \cite{ST}.

\begin{equation} 
\label{eq3}
\begin{split} 
\boldsymbol{x}^{opt} = \argmin_{\substack{x}}  f_{pred} (\boldsymbol{x}).
\end{split} 
\end{equation}

Here, the result of calculation based on quantum annealer is explained.
First, to confirm the accuracy of the calculation based on quantum annealer, the difference between the optimized cost function value $E^{opt}=min f_{pred}(x)$ 
and the value $E^{*} = \sum Q_{ij} x_i^{opt} x_j^{opt} $, in which calculated back with the optimized fingerprint $x_i^{opt}$ and the coefficient of the prediction model $Q_{ij}$ in Eq.(\ref{diff}) is calculated.
\begin{equation} 
\label{diff}
\begin{split} 
\delta = \frac{| E^{opt} -E^{*} |}{E^{opt}} = \frac{| E^{opt} - \sum Q_{ij} x_i^{opt} x^{opt}_j |}{E^{opt}}.
\end{split} 
\end{equation}
As a result, the calculated difference $\delta$ is $7.51 \times 10^{-15}$ (i.e. eleven-nine accuracy).
Second, the optimized cost function value $E^{opt}$ is compared to the minimum value in train data, which is given as $9.99 \times 10^{-5}$.
The sweep time dependence of the optimized value is calculated.
When the number of sweep time is 500, the optimized value is given as $1.40 \times 10^{-5}$, which is lower than the minimum value in train data.
The optimized fingerprint $x^{opt}_i$
is not included in train data, which is confirmed by calculating Hamming distance between them.  
For sweep time 100, on the other hand, the optimized value is given as $ 1.63 \times 10^{-4}$, which is bigger than the minimum value in train data.

\begin{figure}[htbp]
\includegraphics[width=.75\linewidth]{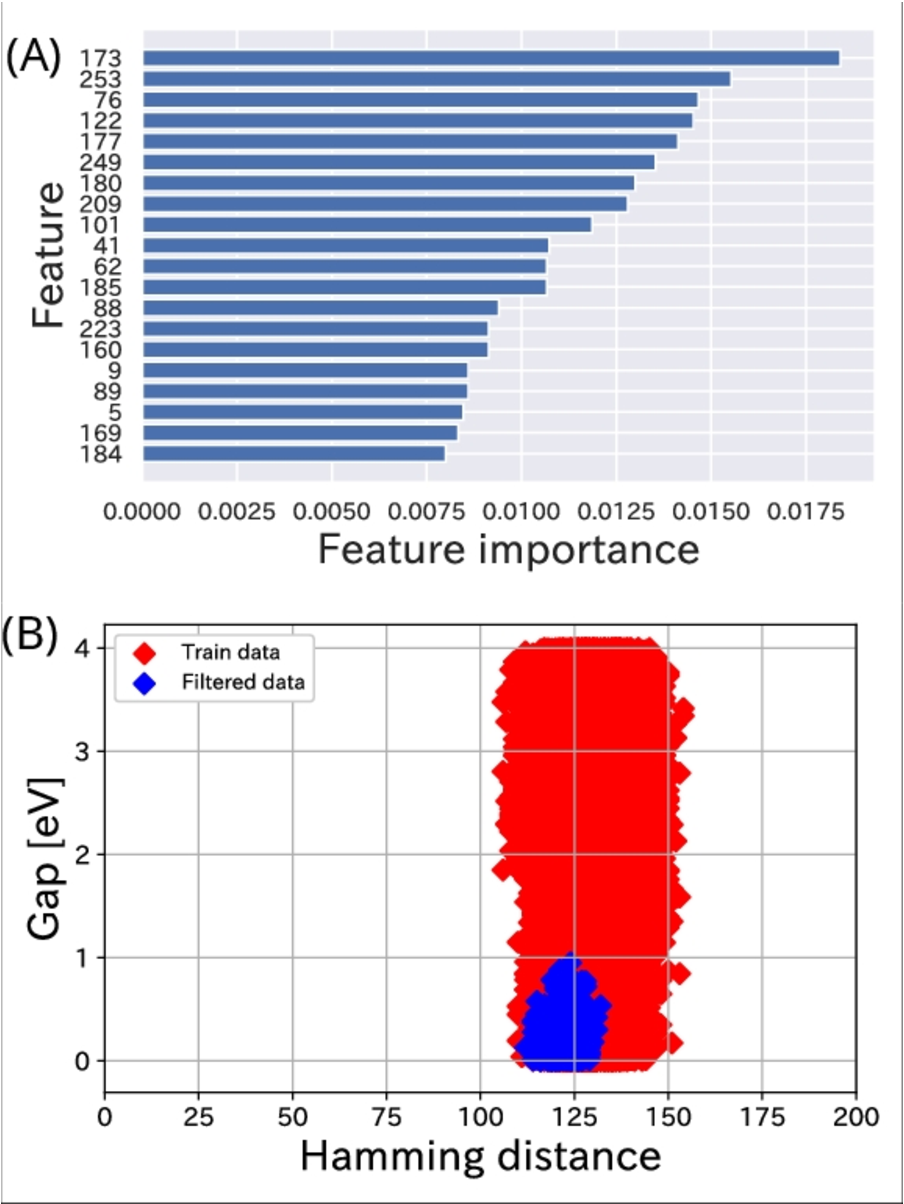}
\vspace*{-4.0cm}
\caption{
(Color online)
(A)The top 20 of feature importance of 1000 annealing sampling data calculated based on light BGM model.
(B)Filtering the 110000 train data by 20 fingerprints extracted from the above analysis,
600 molecular structures are extracted from the train data  data. 
This result suggests that the researching range is screened to 0.5$\%$ (=600/110000),
which indicates the effect of virtual screening based on quantum annealing.
}
\label{f4}
\end{figure}

As a property of quantum annealing, sampling follows the Gibbs-Boltzmann distribution.\cite{Sampling_LosAlamos}
This property of Gibbs-Boltzmann distribution is expected to be utilized in machine learning.
In this study, Gibbs-Boltzmann sampling data is made with openjij.
The feature importance of each fingerprint was evaluated from the frequencies obtained by 1000 sampling data
using LightGBM model (Gradient Boosting Decision Tree).
From the above result, top 20 of feature importance is selected as shown in Fig.\ref{f4} (A).
To judge whether these 20 feature importance are used as 0 or 1, the optimized fingerprint $x_i^{opt}$ are referenced.
Finally, filtering the 110000 train data by 20 feature importance, 600 molecular structures were extracted,
which has small distance from target HOMO-LUMO gap $\Delta^{\ast}$ given in Eq.(\ref{eq0}).
This result suggests that the researching range is screened to 0.5$\%$ (=600/110000) as shown in Fig. \ref{f4} (B),
which indicates the effect of virtual screening based on quantum annealing.

We stuided on extracting feature importance from sampling data generated by Annealer and studied chemical space reduction.
In this paper, feature importance was extracted from sampling data generated by quantum Annealer, and the possibility of reducing chemical space was studied.
Based on QM9 database, SMILES is converted to fingerprint which corresponds to 0,1 state in Ising model.
SGD regressor, only the linear term of quadratic term, is used to predict target HOMO-LUMO gap.
Note that the coefficient of the quadratic term is called QUBO which is used for annealing calculation to minimize the cost function.
The cost function is given as the distance from the target value of HOMO-LUMO Gap.
The feature importance of the sampling data is exctracted, which makes it possible  to screen the whole search space to 0.5 \%.
This result suggests the acceleration of material research with virtual screening.

This is the first application that uses quantum annealing to extract material feature importance, and it has proven to be effective.
In order to verify the effectiveness of this method, a comparison with existing methods is being conducted on actual materials.
The results of the comparison and verification will be published in the near future.

\begin{acknowledgment}
The authors gratefully acknowledge 
Dr. Yoshida, president of the LG Japan Lab Inc.
for helpful discussions and providing an environment suitable for our research. 

\end{acknowledgment}

\clearpage
\begin{equation}
\begin{split}
\\ \nonumber \\%
\\ \nonumber \\%
\\ \nonumber \\%
\\ \nonumber \\%
\\ \nonumber \\%
\end{split}
\end{equation}

\begin{equation}
\begin{split}
\\ \nonumber \\%
\\ \nonumber \\%
\\ \nonumber \\%
\end{split}
\end{equation}

\begin{equation} 
\begin{split}
\\ \nonumber \\%
\\ \nonumber \\%
\end{split}
\end{equation}

\end{document}